\documentclass[smallextended]{svjour3}       
\smartqed  
\usepackage{graphicx}
\usepackage{amsmath}
\usepackage{amsfonts}
\usepackage{amssymb}
\newcommand{\sph}[1]{\left\langle #1 \right\rangle}

\begin{document}

\title{Smoothed Particle Hydrodynamics modeling of granular column collapse}

\titlerunning{SPH modeling of granular column collapse}        

\author{Kamil Szewc}


\institute{K. Szewc \at
              Institute of Fluid-Flow Machinery \\
              Polish Academy of Sciences \\
              ul. Fiszera 14 \\
              80-231 Gda\,nsk \\
              \email{kszewc@imp.gda.pl}           
}

\date{Received: date / Accepted: date}

\maketitle

\begin{abstract}
The Smoothed Particle Hydrodynamics (SPH) is a particle-based, Lagrangian method
for fluid-flow simulations. In this work, fundamental concepts of this method are
first briefly recalled. Then, the ability to accurately model granular materials 
using an introduced visco-plastic constitutive rheological model is studied.
For this purpose sets of numerical calculations (2D and 3D) of the fundamental
problem of the collapse of initially vertical cylinders of granular materials
are performed. The results of modeling of columns with different aspect ratios
and different angles of internal friction are presented. The numerical outcomes are 
assessed not only with respect to the reference experimental data but also with 
respect to other numerical methods, namely the Distinct Element Method
and the Finite Element Method. In order to improve the numerical efficiency
of the method, the Graphics Processing Units implementation is presented and 
some related issues are discussed. It is believed that this study corresponds to a new application of SPH approaches for simulations of granular media and results
reveal the interest of this method to capture fine details of processes of
such complex problems as waves-seabed interactions. 
\end{abstract}

\section{Introduction}
Problems involving large granular media deformations are active research in 
the fields of geomechanics and natural hazard management. Particular attention is paid
to understand the processes and to learn how to predict the run-outs of
rock and debris avalanches or landslides which can be very destructive.
Therefore, the accurate modeling of such flows is invaluable.

In order to simulate flowing granular media, many researchers derived the semi-empirical
depth-averaged models, see~\cite{Savage and Hutter 1989,Anderson and Jackson 1992,Poliquen 1999,Iverson and Denlinger 2001,Lajeunesse et al. 2005}.
The main disadvantage of such models is inability to accurately model
high but narrow stacks of granular materials. This drawback implied a need
to develop more detailed computational models for granular material dynamics.
One of the most widely used methods to simulate such problems
is the Distinct Element Method (DEM)~\cite{Cundall and Strack 1979}.
The first application of DEM to model granular flows was proposed 
in~\cite{Cleary and Campbell 1993}. Further improvements have been described
in~\cite{Staron and Hinch 2005,Staron and Hinch 2007,Tang et al. 2009,Zenit 2005,Utili et al. 2015}.
Nowadays, the DEM method is widely accepted as an effective approach for modeling
granular and discontinuous materials. Even through the DEM approach is computationally
expensive, this approach is easy to be written in a parallel manner.
Despite many advantages of DEM, there is one relevant disadvantage --
this approach is designed to discrete materials modeling, therefore the introduction
of a new physics can be very complex. For example, to model interactions of granular material
with fluid-flow, the coupling with another CFD method is necessary.
One of the alternative methods which allow to deal with multi-physics is the
Finite Element Method (FEM), see~\cite{Zienkiewicz et al. 2005} for details.
To the best of our knowledge, the first application of FEM to model the granular 
material collapse was proposed in~\cite{Crosta et al. 2009}. The authors used 
the continuum approach based on an elasto-plastic model.
The main disadvantage of FEM is the grid-based nature of this method.
When large deformations occur, the FEM approach suffers from grid distortions.
However, in the last decade, many so-called meshless methods have been developed. 

One of the most mature and commonly used approach is 
the Smoothed Particle Hydrodynamics method (SPH). In the early stage it was developed to simulate some astrophysical phenomena at the hydrodynamic level~\cite{Monaghan 1992}. The main idea behind the SPH method is to introduce kernel interpolants for flow quantities in order to represent fluid dynamics by a set of particle evolution equations. Due to its Lagrangian nature, for multiphase flows, there is no necessity to handle (reconstruct or track) the interface shape as in the grid-based methods. Therefore, there is no additional numerical diffusion related to the interface handling. For this reason and the fact that the SPH approach is well suited to problems with large density differences, free-surfaces and complex geometries, 
the SPH method is increasingly used for hydro-engineering and geophysical applications, 
for review see~\cite{Monaghan 2012,Violeau 2012}.
The first attempts to model the granular materials using the SPH approach was presented
by \emph{Bui et al. (2008)}~\cite{Bui et al. 2008,Bui et al. 2014}. The authors decided to use an elasto-plastic (Drucker-Prager) constitutive model. Despite good results compared
with the experimental and numerical data, authors reported a serious tensile instability
problems. 
Another constitutive model was proposed by \emph{Ulrich et al. 2013}~\cite{Ulrich et al. 2013},
where the granular material is treated as a fluid with a variable viscosity 
(a visco-plastic rheological model), however, authors have not presented any validations
of this approach. 

In the present work, the ability to model granular materials using the SPH method
and the visco-plastic model is studied.
For this purpose, it was decided to perform a set of numerical calculations 
(in 2D and 3D) of the fundamental problem of the collapse of initially vertical cylinders
of granular materials. The obtained results were compared with other numerical (DEM and FEM)
and the experimental data.
On of the drawbacks of the mesh-free methods is much lower numerical efficiency compared
with the grid-based approaches. However, similarly to the DEM method, the SPH approach
numerical implementations present a high degree of spatial data locality and significant
number of independent calculations, therefore the code can be easily written in
a massively parallel manner. In recent years new techniques allowing numerical 
simulations to be performed using Graphics Processing Units (GPU) have been developed.
The massive parallel capability of modern GPUs allows simulations of large systems
to be performed using cheap desktop computers.
For the purpose of this study, it was decided to implement the SPH method using 
GPU programming techniques. Some issues related to the GPU computations are discussed.

The paper is organized as follows: in Section~\ref{sec:sph formulation} the brief
introduction to SPH is presented; in Section~\ref{sec:granular material modeling} 
the visco-plastic rheological model of granular materials is introduced;
in Section~\ref{sec:graphics processor unit implementation} the implementation
on GPU is discussed; then in Section~\ref{sec:numerical results} the obtained
numerical results are presented. In order to validate the model, the following criteria are
taking into account: the granular deposit evolution (Section~\ref{sec:shape evolution}),
run-out distances (Section \ref{sec:run-out distances}), the energy contribution 
(Section \ref{sec:energy contribution}) and the inclination of the failure
plane (Section~\ref{sec:inclination of failure plane}). The numerical efficiency
is discussed in Section~\ref{sec:numerical efficiency}.

\section{SPH formulation} \label{sec:sph formulation}
The full set of governing equations for incompressible viscous flows is composed of the Navier-Stokes (N-S) equation
\begin{equation}
    \frac{d\mathbf u}{dt} = -\frac{1}{\varrho} \nabla p + \frac{1}{\varrho} \left(\nabla \mu \cdot \nabla \right) \mathbf u + \mathbf g,
\end{equation}
where $\varrho$ is the density, $\mathbf u$ velocity, $t$ time, $p$ pressure, $\mu$ the dynamic viscosity and $\mathbf g$ an acceleration (gravity in this work); 
the continuity equation
\begin{equation}
    \frac{d\varrho}{dt} = -\varrho\nabla \cdot \mathbf u \xrightarrow{\varrho=const} \nabla \cdot \mathbf u = 0,
\end{equation}
and the advection equation (Lagrangian formalism)
\begin{equation}
    \frac{d\mathbf r}{dt} = \mathbf u,
\end{equation}
where $\mathbf r$ denotes position of the fluid element.

The governing equations can be expressed in the SPH formalism in many different ways. In general, two SPH approximations: integral interpolation and discretization, lead to the the basic SPH relation
\begin{equation} \label{basic SPH}
    \sph{A}(\mathbf r) = \sum_b A(\mathbf r_b) W(\mathbf r - \mathbf r_b, h) \Omega_b,
\end{equation}
where $A$ is a physical field (for the sake of simplicity we consider a scalar field only), $W$ is a weighting function (kernel) with parameter $h$ called the smoothing length, while $\Omega$ is the volume of the SPH particle. There are many possibilities for the choice of 
$W(\mathbf r, h)$. The kernel shape is the main reason for the appearance of the tensile
instability resulting in particle clumping~\cite{Swegle et al. 1995} -- the process
from which the results in \cite{Bui et al. 2008} suffer. 
In \cite{Szewc et al. 2012a}
the authors performed series of fluid-flow simulations and
showed that using the Wendland kernel~\cite{Wendland 1995} in the form
\begin{equation} \label{quintic Wendland}
W(\mathbf r, h) = C  \left \{
 \begin{array}{cl}
  \left( 1-q/2 \right)^4 (2q+1) &  \text{for} \;\; q \leq 2, \\
  0                                     &  \text{otherwise},     \\
  \end{array}
  \right.
\end{equation}
where $q = |\mathbf r|/h$ and the normalization constant is $C=7/4\pi h^2$ (in 2D) 
or $21 / 16\pi h^3$ (in 3D), the tensile instability does not appear.
Therefore, in this work, we decided to use the kernel in the form of Eq.~(\ref{quintic Wendland}).
For more details how the choice of the kernel and the smoothing length 
affect results 
see~\cite{Szewc et al. 2012a}. It is important to note here that the SPH basic approximation, Eq.~(\ref{basic SPH}), is common also in other numerical particles-based approaches, e.g. Moving Particle Semi-implicit Method (MPS)~\cite{Koshizuka et al. 1998}. The SPH method differs from other methods in aspect of approximation of differentiation operator. Assuming the kernel symmetry, nabla operator can be shifted from the action on the physical field to the kernel
\begin{equation}
    \sph{\nabla A}(\mathbf r) = \sum_b A(\mathbf r_b) \nabla W(\mathbf r - \mathbf r_b, h) \Omega_b.
\end{equation}

It is important to note that although different SPH formulations can be obtained from the same governing equations, some of them may not by applicable for certain types of flows,
see~\cite{Szewc et al. 2012b}.
One of the most common SPH form, enabling accurate calculations in the widest
number of types of flow, is the N-S pressure term proposed by
\emph{Colagrossi and Landrini (2003)}~\cite{Colagrossi and Landrini 2003}
\begin{equation} \label{pressure term}
    \sph{\frac{\nabla p}{\varrho}}_a = -\sum_b m_b \frac{p_a + p_b}{\varrho_a \varrho_b}  \nabla_a W_{ab},
\end{equation}
where $\nabla_a W_{ab}=\nabla_a W(\mathbf r_a - \mathbf r_b, h)$.
In the present work, we decided to perform calculations of pressure term
using this variant.
The corresponding (variationally consistent~\cite{Grenier et al. 2009,Monaghan 2005}) 
continuity equation takes a form
\begin{equation}
	\sph{\frac{d\varrho}{dt}}_a = \varrho_a \sum_b \frac{m_b}{\varrho_b} \mathbf u_{ab} \cdot \nabla_a W_{ab},
\end{equation}
where $\mathbf u_{ab} = \mathbf u_{a} - \mathbf u_{b}$.
The viscous N-S term, because of the efficiency requirements, is expressed as
a combination of the finite difference and the SPH approach 
(as in the MPS approach~\cite{Koshizuka et al. 1998})
\begin{equation}
	\sph{\frac{1}{\varrho} \left(\nabla \mu \cdot \nabla \right) \mathbf u}_a 
	= \sum_b m_b \frac{\mu_a + \mu_b}{\varrho_a \varrho_b}
	\frac{\mathbf r_{ab} \cdot \nabla_a W_{ab}(h)}{r_{ab}^2 + \eta^2}
	\mathbf u_{ab},
\end{equation}
where $\eta=0.01h$ is a small regularizing parameter used to avoid \emph{NaN}s
when divide by the numerical zero.
Because SPH is a Lagrangian approach, the particle advection equation completes
the system
\begin{equation}
	\frac{d \mathbf r_a}{dt} = \mathbf u_a.
\end{equation}

In the present work, we decided to use the most common method of implementing the incompresibility -- Weakly Compressible SPH (WCSPH). It involves the set of governing equations closed by a suitably-chosen, artificial equation of state, $p=p(\varrho)$. Following the mainstream, we decided to use the Tait's equation of state
\begin{equation}
    p = \frac{c^2 \varrho_0}{\gamma} \left[\left( \frac{\varrho}{\varrho_0} \right)^\gamma - 1 \right],
\end{equation}
where $\varrho_0$ is the initial density. The sound speed $c$ and a parameter $\gamma$ are suitably chosen to reduce the density fluctuations down to $1\%$. In the present work we set $\gamma=7$ and $c$ at the level at least 10 times higher than the maximal fluid velocity. It is worth noting that two alternative incompressibility treatments exists: Incompressible SPH (ISPH) where the incompressibility constraint is explicitly enforced though the pressure correction procedure to satisfy $\nabla \cdot \mathbf u=0$~\cite{Szewc et al. 2012a,Cummins and Rudman 1999,Hu and Adams 2007,Shao and Lo 2003,Xu et al. 2009} and Godunov SPH (GSPH) where the acoustic Riemann solver is used~\cite{Rafiee et al. 2012}.
In the present work, the boundary conditions are fulfilled applying 
the ghost-particle method~\cite{Szewc et al. 2012a,Cummins and Rudman 1999}.

To assure the stability of the SPH scheme several time step criteria
must be satisfied:
\begin{equation} \label{cfl}
	\delta t < 0.125 \frac{h}{c + u_{\text{max}}}, \quad 
	\delta t < 0.125 \frac{\varrho h^2}{\mu_{\text{max}}}, \quad
	\delta t < 0.125 \left( \frac{h}{g} \right)^{\frac{1}{2}},
\end{equation}
where $u_\text{max}$ and $\mu_{\text{max}}$ are respectively the maximal
particle velocity and the maximal particle viscosity in the domain.

\section{Granular material modeling} \label{sec:granular material modeling}

In order to simulate the granular materials using the SPH approach, we decided to 
adopt the visco-plastic rheological model first used in SPH by 
\emph{Ulrich et al. (2013)}~\cite{Ulrich et al. 2013}. 
Equations that predict the shape of the general flow curve need at least four 
independent parameters. A common example is the \emph{Cross (1965)}~\cite{Cross 1965}
equation
\begin{equation} \label{cross}
  \frac{\mu_0 - \mu}{\mu-\mu_{\infty}} = (K \dot \gamma)^m,
\end{equation}
where $\mu_0$ and $\mu_{\infty}$ refer to the asymptotic values of viscosity,
while $K$ and $m$ are constant.
The shear strain rate, $\dot \gamma$, can be defined as
\begin{equation}
  \dot \gamma = \sqrt{2\dot \epsilon^{ij} \dot \epsilon^{ij}},
\end{equation}
where
\begin{equation}
  \epsilon^{ij} = \frac{1}{2}\left(\frac{\partial u^i}{\partial x^j} + \frac{\partial u^j}{\partial x^i} \right).
\end{equation}
In the case of the granular phase, $\mu_0$ corresponds to the viscosity of solid
(low values of $\dot \gamma$, elastic limit), while $\mu_{\infty}$ is the viscosity of 
grains above elastic limit (high values of $\dot \gamma$).
Therefore, we may assume that $\mu \ll \mu_0$, which reduces Eq.~(\ref{cross}) to the
\emph{Sisko (1958)}~\cite{Sisko 1958} model
\begin{equation}
  \mu = \mu_{\infty} + \frac{\mu_0}{(K \dot \gamma)^m} = \mu_{\infty} + K_2 \dot \gamma^{n-1}.
\end{equation}
Assuming $n=0$, we get
\begin{equation} \label{bingham}
  \mu = \mu_{\infty} + \frac{K_2}{\dot \gamma},
\end{equation}
which is commonly known as the \emph{Bingham (1916)} model~\cite{Bingham 1916}.
With some simple redefinition of parameters Eq.~(\ref{bingham}) can be written as
\begin{equation}
  \tau = \mu_0 \dot \gamma + \tau_{\text{yield}},
\end{equation}
where $\tau$ is the shear stress, $\tau_{\text{yield}}$ is the yield stress.
In this model, the material behaves as a solid body until the shear stress exceeds
the yields stress (reaching the critical state) and large deformations may occur. 
One of the commonly used models is the Mohr-Coulomb failure criterion~\cite{Coulumb 1776},
in which the shear strength of soil is expressed as a combination of adhesion
and friction components
\begin{equation}
  \tau_{\text{yield}} = c + \sigma_n \tan \varphi,
\end{equation}
where $c$ is the cohesion, $\varphi$ is the internal friction angle, while 
$\sigma_n$ is the normal stress. 
However, it is important to note that $c$ and $\varphi$ are not fundamental
properties of material. Both depend on the effective stress~\cite{Lambe and Whitman 1991}.
However, for the purposes of the present study, it is sufficiently to assume 
that $c$ and $\varphi$ are fundamental material constants.

Assuming that $\sigma_n = p$, the final form of the granular material model
takes the form
\begin{equation} \label{rheological}
  \mu =
  \begin{cases}
    \mu_{\infty} + (c+p\tan \phi)/\dot \gamma,       & \mu < \mu_{\text{solid}}, \\
    \mu_{\text{solid}},       & \mu \geq \mu_{\text{solid}}, \\
  \end{cases}
\end{equation}
where $\mu_{\text{solid}}$ is introduced, due to numerical efficiency reasons, 
to avoid extremely high values of viscosity, which may lead to extremely 
small time steps (due to CFL).

\section{Graphics Processor Unit implementation} \label{sec:graphics processor unit implementation}

\begin{figure}
  \centering
  \includegraphics[width=0.99\textwidth]{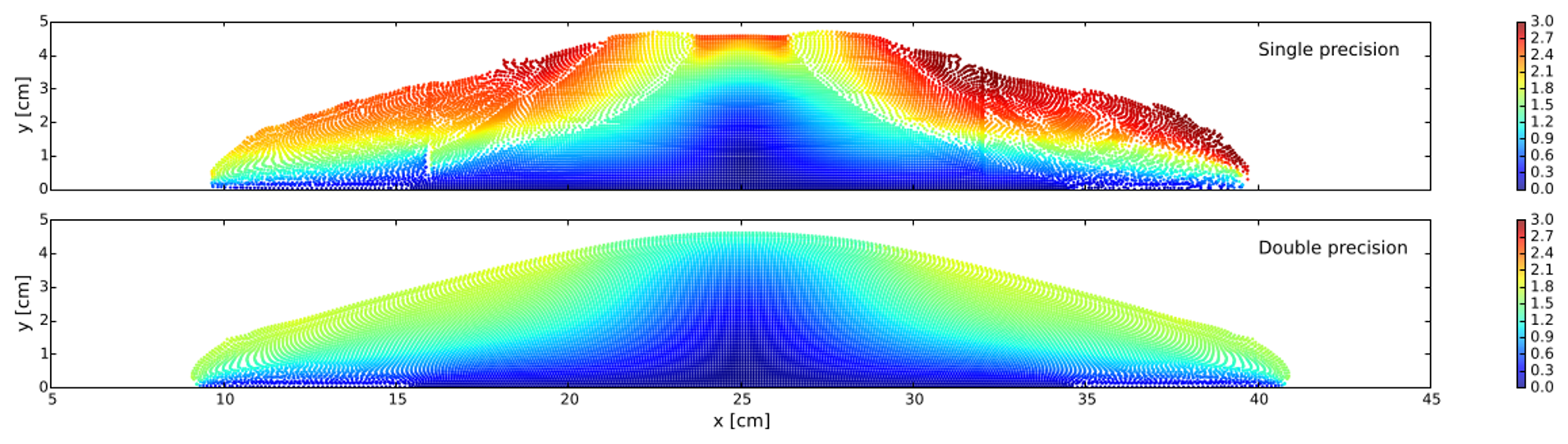}
  \caption{2D granular collapse velocity fields [cm s$^{-1}$] at $t=350$ 
  ($a=0.55$ -- for details, see Sect.~\ref{sec:introduction});
  the SPH simulations obtained with: (a) single precision and
  (b) double precision floating point numbers.}
  \label{fig:error}
\end{figure}

The modern desktop CPUs, such as \emph{Intel i7-4790K},
have 4 physical cores (8 virtual cores via hyper-threading) 
with the base frequency about $4$ GHz.
For comparison, the modern desktop GPUs, such as 
\emph{Nvidia GeForce GTX 980}, have more than $2\cdot 10^3$ cores
with the base frequency about $1$ GHz.
Therefore, the advantage of using GPU accelerators for HPC is obvious.
The GPU cards were designed to accelerate the creation of images
in a frame buffer to stream them onto a display, therefore the double
precision was not needed for such a task. 
Due to this, most of the desktop GPUs are built to support mainly the single
precision calculations. 
There is a possibility to run tasks in double precision, 
but, it results in a significant drop of performance
(officially, for \emph{Nvidia Maxwell series} it is about 16 times).
It is important to note that for the most applications of the SPH approach
the numerical errors related to the used approximations are much higher
than the truncation errors, therefore, many researchers decided to perform
the SPH calculations using GPUs with the single precision, 
see~\cite{Herault et al. 2010,Szewc 2014,Rustico et al. 2014,Crespo et al. 2015}.
However, since in our case the kinematic viscosity of a granular material can change the value more then $5$ orders of magnitude during a simulation, the single precision is 
not enough. The influence of the floating point number precision on the results
is presented in Fig.~\ref{fig:error}. 

The problem of the double precision floating point numbers 
in the SPH modeling on GPU has been recently discussed
in~\cite{Herault et al. 2014} and \cite{Dominguez et al. 2014}.
To avoid this problem the authors proposed to use such techniques as
the cell relative coordinates (to avoid problems in domains of high aspect ratios)
or the compensated algorithms like Kahan sum (to sum over large numbers of values).
Unfortunately, none of the proposed algorithms could correct the problem
of strongly inhomogeneous viscosity in domain. Therefore, in the present work, 
to avoid inaccuracies, we decided to perform calculations using 
the double precision explicitly.
The influence on the numerical efficiency is discussed in Section~\ref{sec:numerical efficiency}.
For details about the GPU implementation see~\cite{Szewc 2014}.

\section{Numerical results} \label{sec:numerical results}

\subsection{Introduction} \label{sec:introduction}

\begin{figure}
  \centering
  \includegraphics[width=0.49\textwidth]{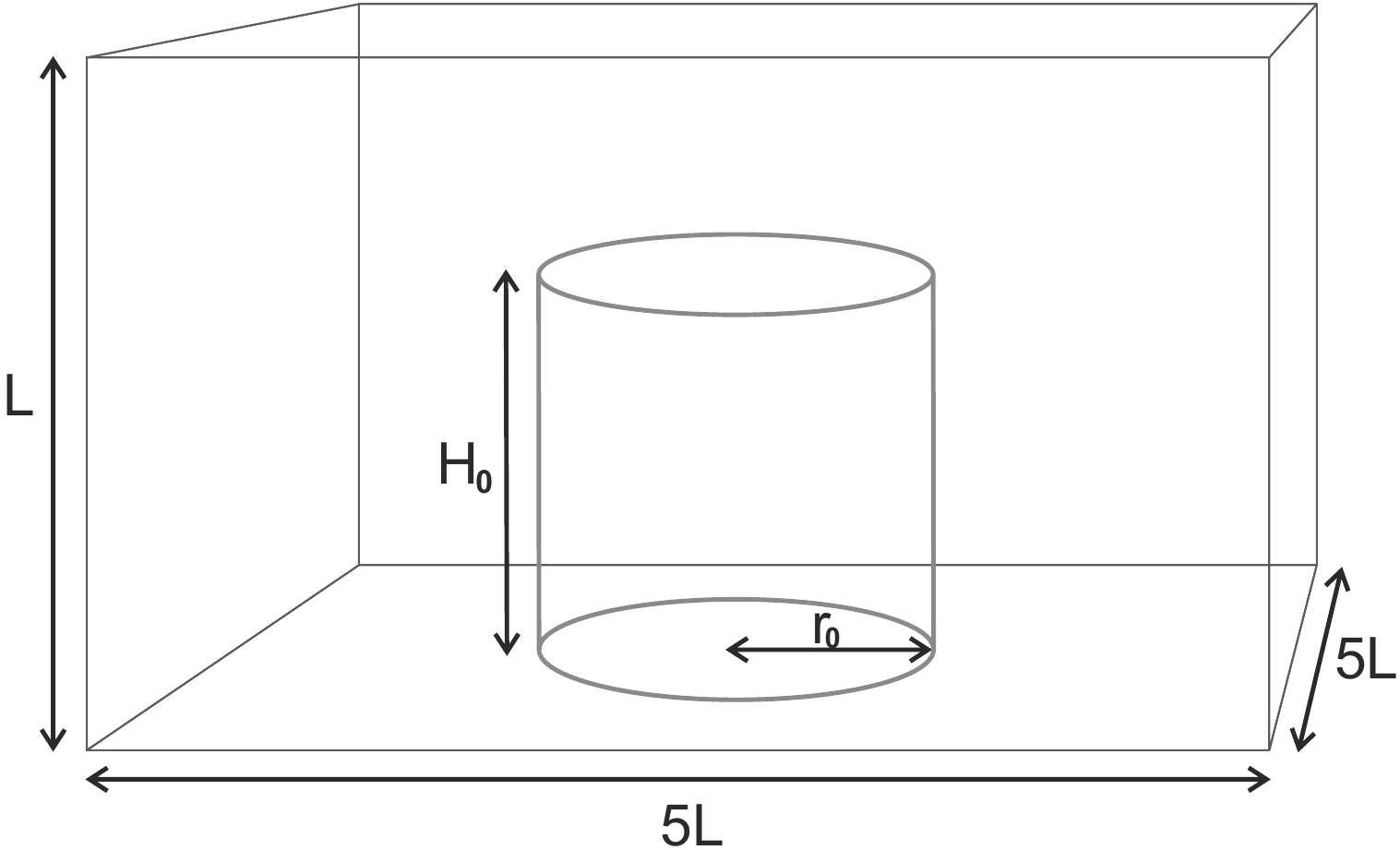}
  \caption{Initial configuration in 3D.}
  \label{fig:configuration}
\end{figure}

The numerical experiments were performed by releasing initially vertical
columns of granular material, see Fig.~\ref{fig:configuration}. 
The initial height, $H_0$, was defined by 
the initial radius $r_0 = 9.7$ cm and the aspect ratio parameter
\begin{equation}
  a = \frac{H_0}{r_0}.
\end{equation} 
The material density $\varrho$ was $2.6$ g cm$^{-3}$. 
The angle of repose $\varphi$ was $30^{\circ}$ 
(except Section ~\ref{sec:inclination of failure plane}). 
The material was chosen as non-cohesive, $c = 0$. 
These are properties of dry sand used in the experiment of 
\emph{Lube et al. (2004)}~\cite{Lube et al. 2004}.
Initially the granular column is placed in the middle of the base of the rectangular
domain of edges $(5L, 5L, L)$ in 3D or $(5L, L)$ in 2D, 
where $L=1.8H_0$ is the domain height. For $a<0.9$ we have chosen $L=10$ cm, 
while for others $L=1.2H_0$.

The SPH simulations were performed for different aspect ratios and 
different numerical resolutions.
In 2D: $a=0$-$10$, $h/L=16$-$64$ and $h/\Delta r=2$.
In 3D: $a=0$-$6$, $h/L=16$-$64$ and $h/\Delta r=1.5625$
(lower than 2D due to the efficiency reasons).
In both cases, we decided to use the Wendland kernel 
(to avoid problems with the tensile instability). 
The speed of sound was $s=1000$ cm s$^{-1}$.
The parameter $\mu_{\infty}$ in Eq.~(\ref{rheological}) was chosen as 
a viscosity of water (at $20^{\circ}C$), $0.01$ g cm$^{-1}$ s$^{-1}$.
The viscosity of solid $\mu_{\text{solid}}$ was chosen as $2000$ g cm$^{-1}$ s$^{-1}$.
Due to the efficiency reasons, see Eq.~(\ref{cfl}), the value of $\mu_{\text{solid}}$
is suitably lowered compared with the real soil.
The side effects of such proceedings are negligible.
The calculations were performed using the double
precision floating point number with the single precision calculations
used only to benchmark the numerical efficiency in Section~\ref{sec:numerical efficiency}.

\subsection{Shape evolution} \label{sec:shape evolution}

\begin{figure}
  \centering
  \includegraphics[width=0.49\textwidth]{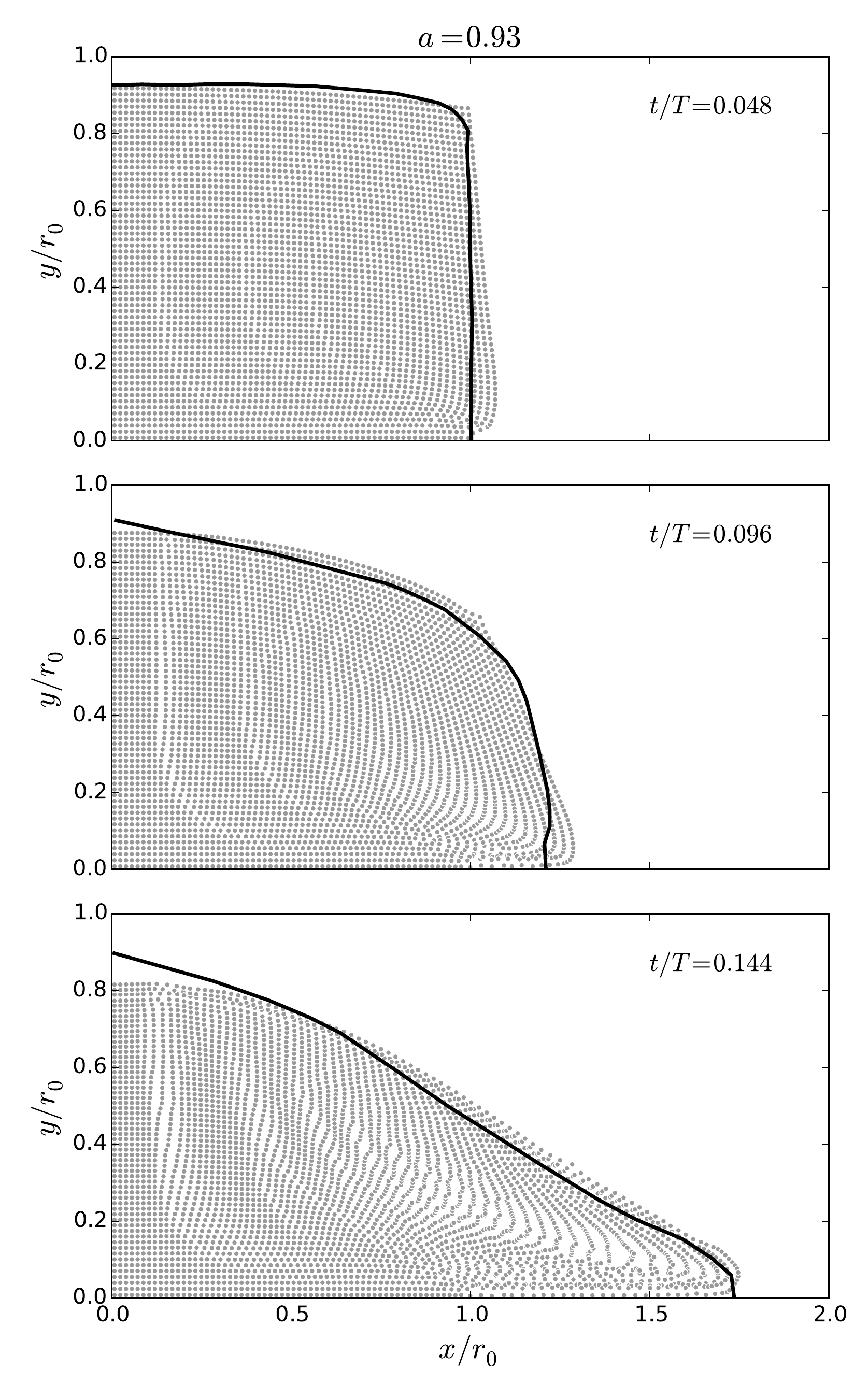}
  \includegraphics[width=0.49\textwidth]{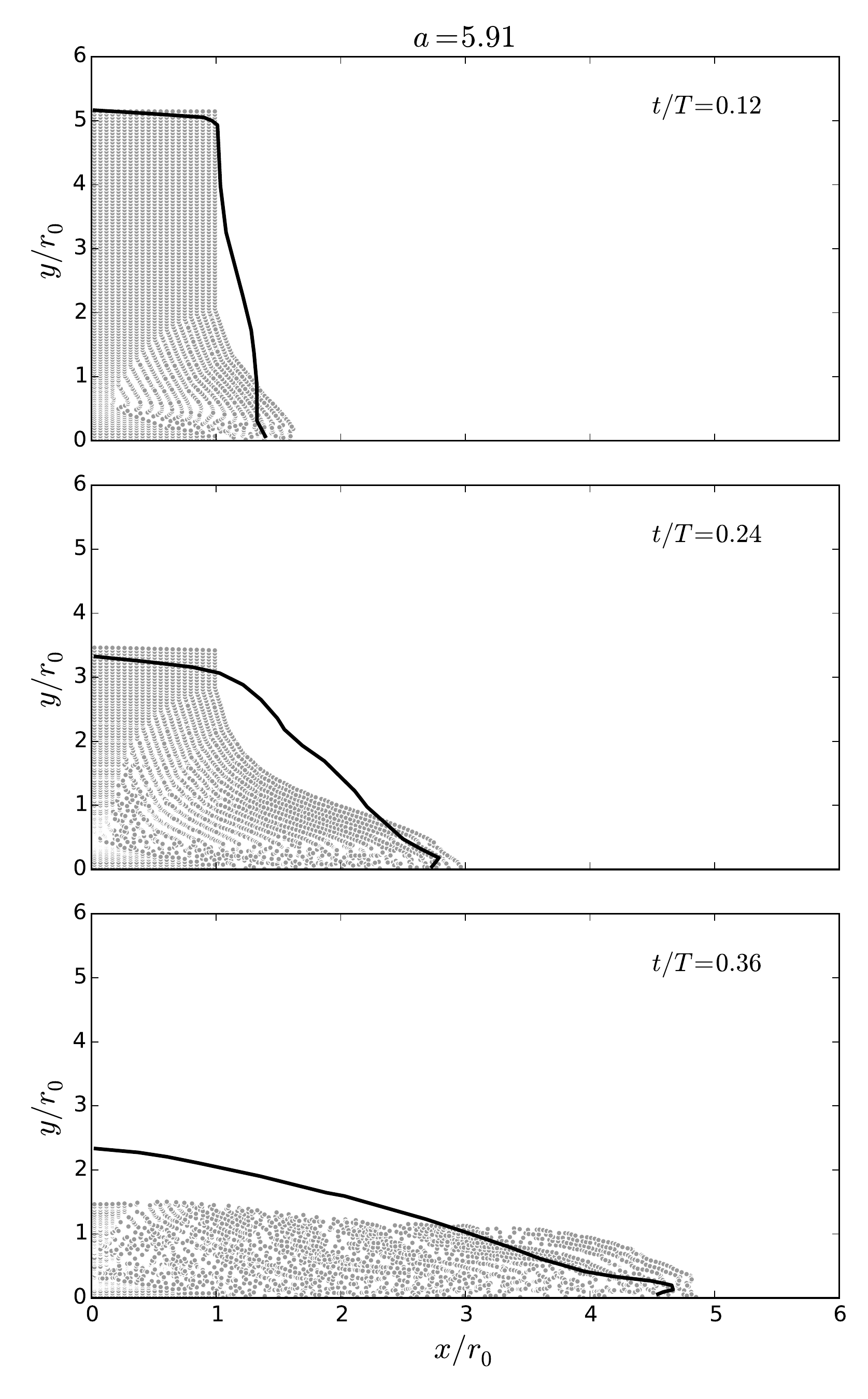}
  \caption{The evaluation of profiles for different initial aspect ratios, $a$, 
  of granular columns (2D);
  the SPH results compared with the DEM reference data \cite{Utili et al. 2015} (solid lines).}
  \label{fig:shapes-2d}
\end{figure}

The simplest technique to check whether the proposed model gives the correct results is 
to compare the calculated profile shapes with other numerical and experimental data. 
In 2D, as a reference data we decided to choose the DEM calculations obtained by
\emph{Utili et al. (2015)}~\cite{Utili et al. 2015}.
The obtained results, for aspect ratios: $a=0.93$ and $a=5.91$, are presented in 
Fig.~\ref{fig:shapes-2d}.
For $a=0.93$ the SPH results show good agreement with the reference data.
In the case of higher aspect ratio, $a=5.91$, the SPH calculations slightly differ 
from the DEM calculations -- mainly the final height of the sample.

\begin{figure}
  \centering
  \begin{tabular}{c|c}
    $a = 0.55$ & $a=2.75$ \\ \hline
    \phantom{o} & \phantom{o} \\
    \includegraphics[width=0.48\textwidth]{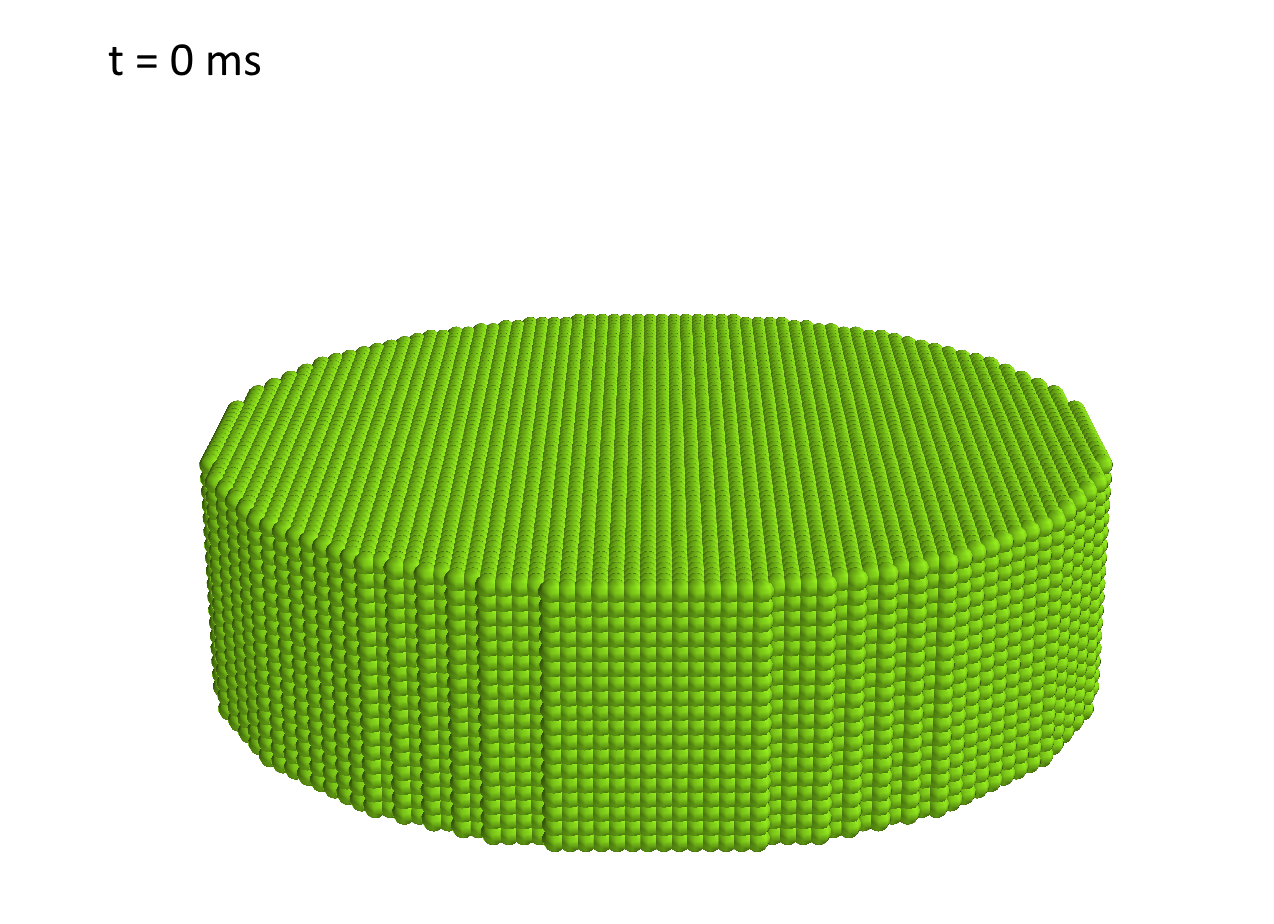} &
    \includegraphics[width=0.48\textwidth]{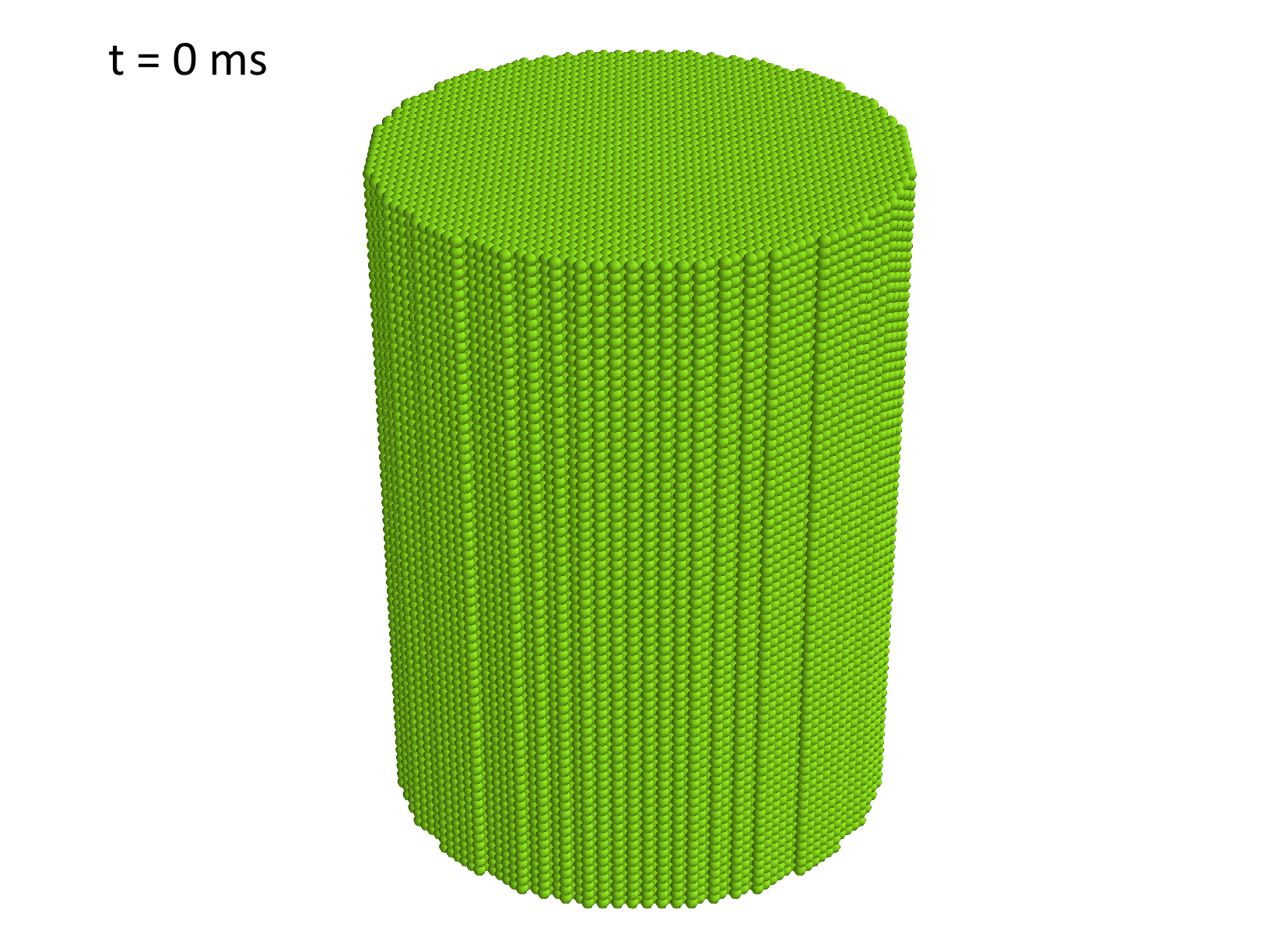} \\
    \includegraphics[width=0.48\textwidth]{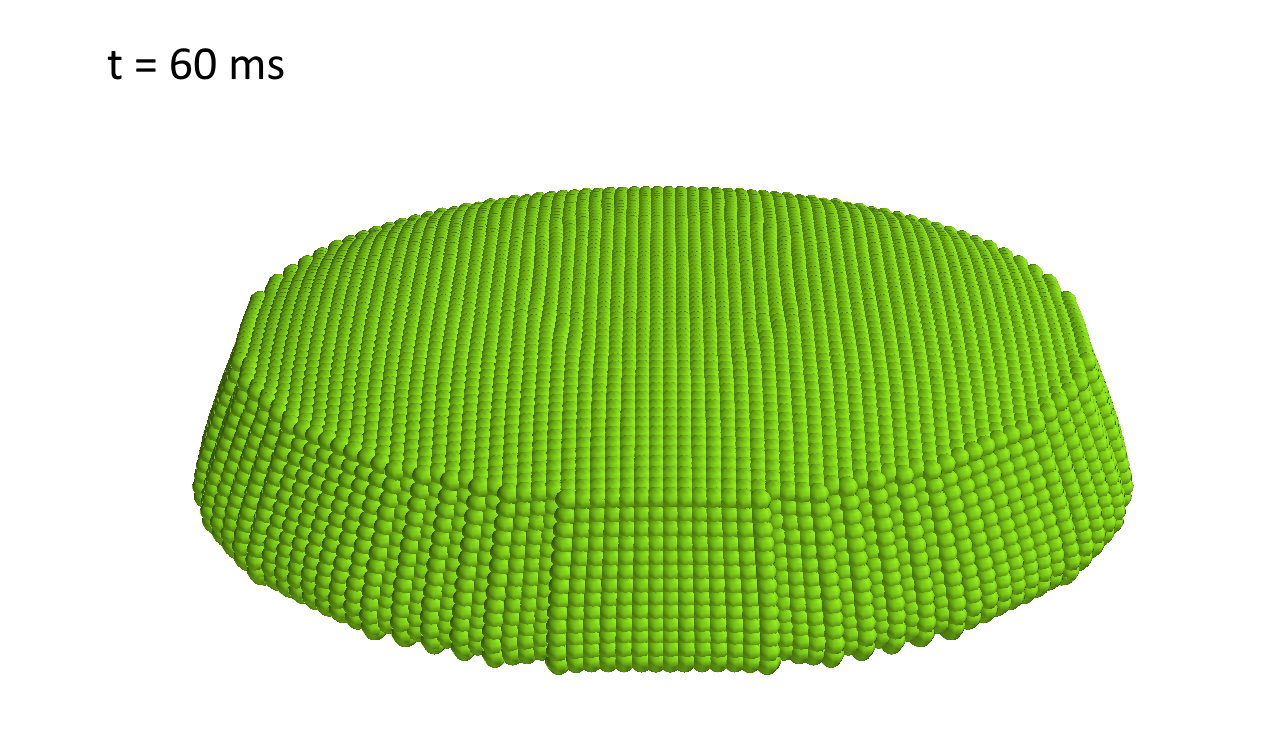} &
    \includegraphics[width=0.48\textwidth]{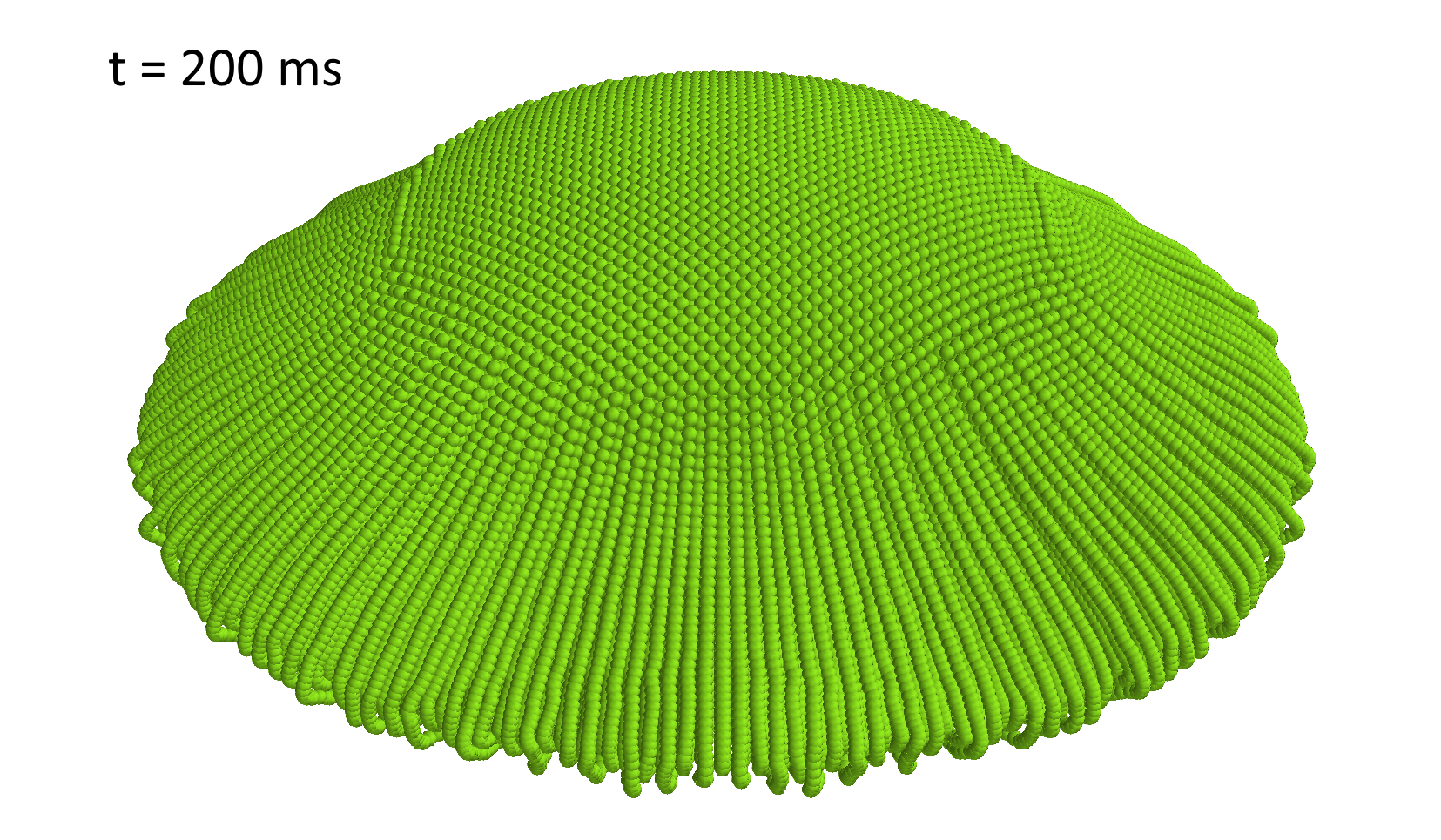} \\
    \includegraphics[width=0.48\textwidth]{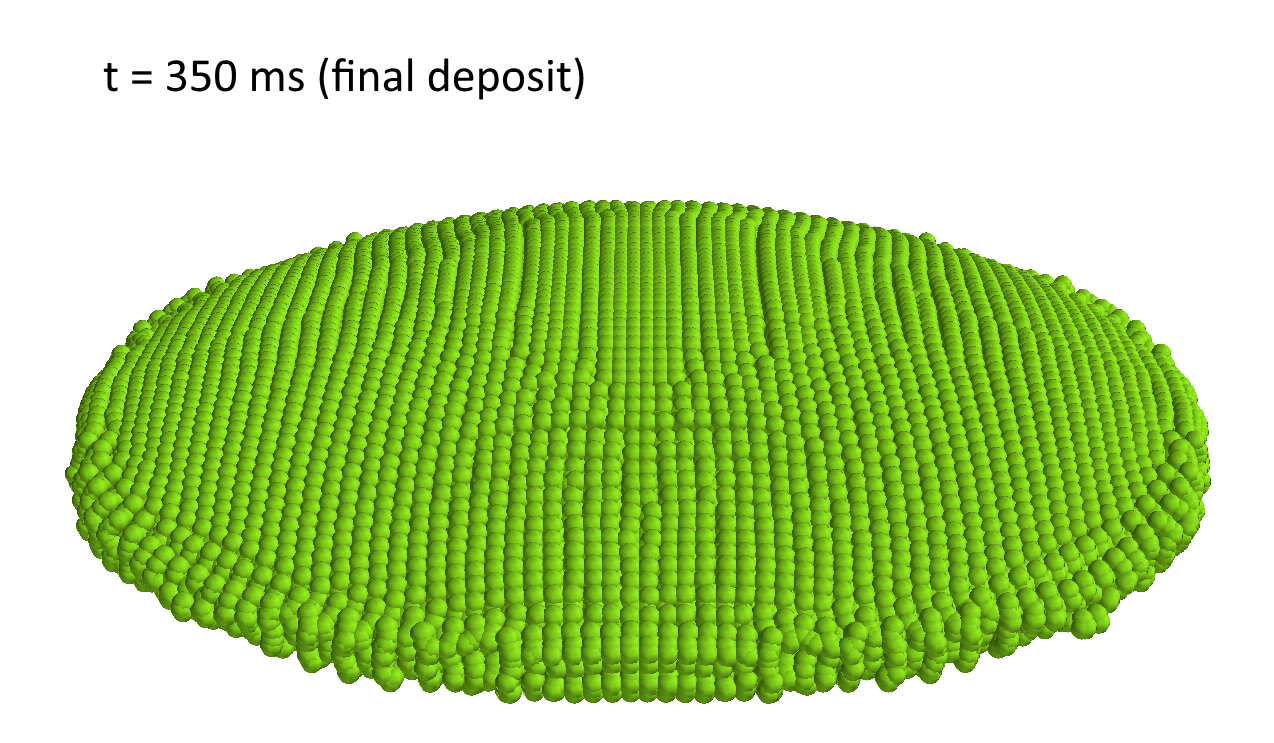} &
    \includegraphics[width=0.48\textwidth]{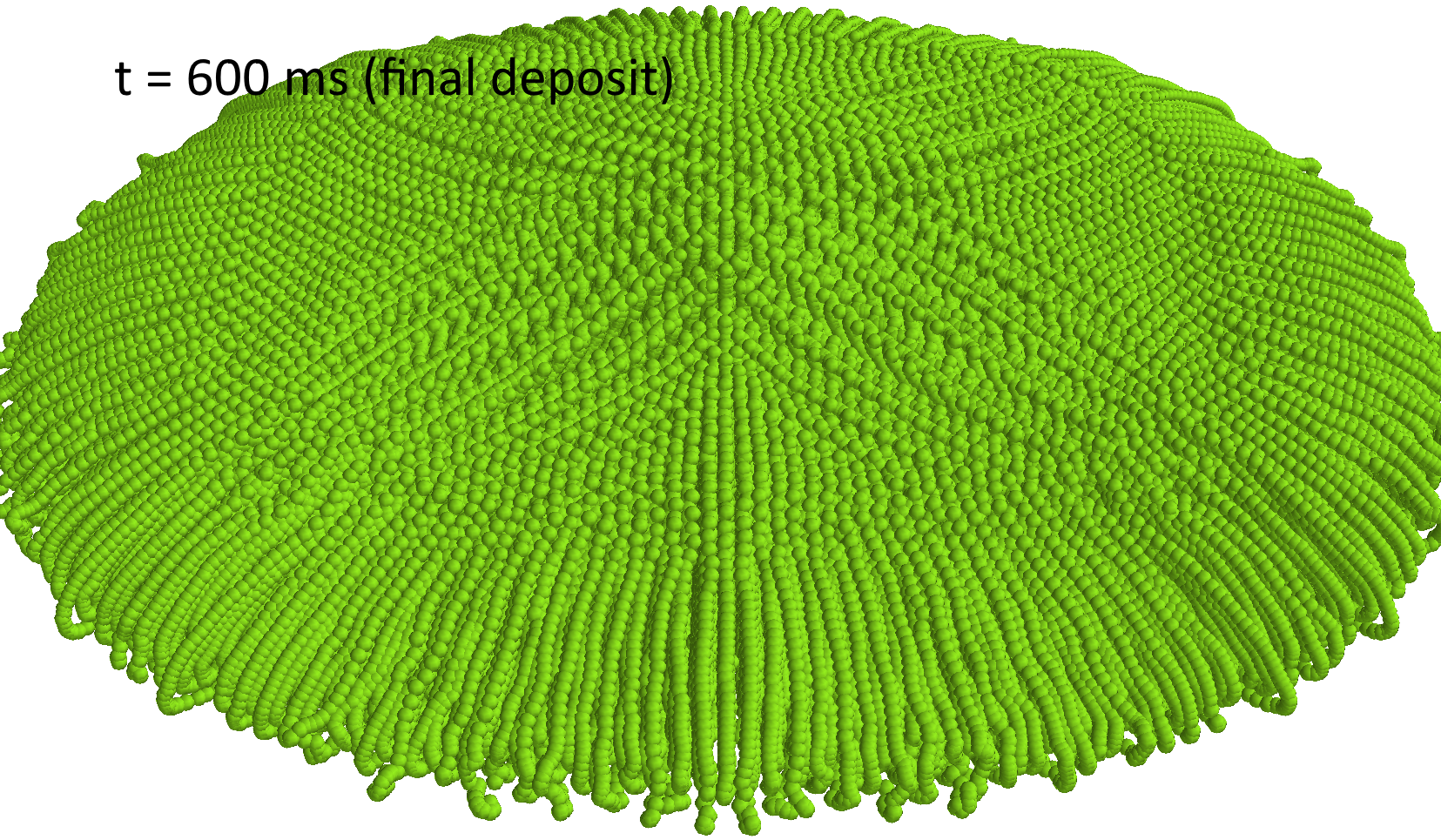} \\
    \includegraphics[width=0.48\textwidth]{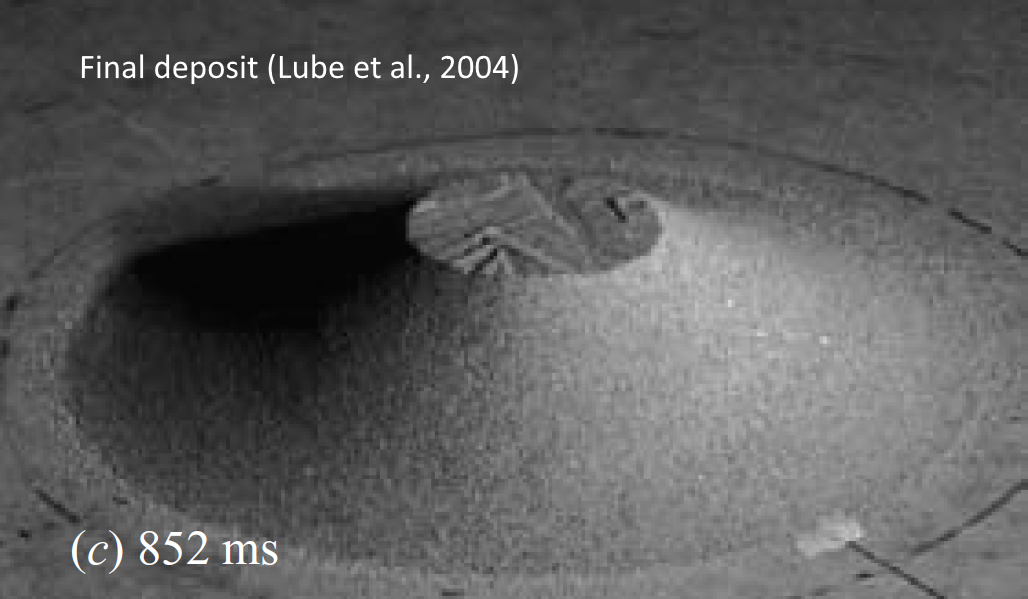} &
    \includegraphics[width=0.48\textwidth]{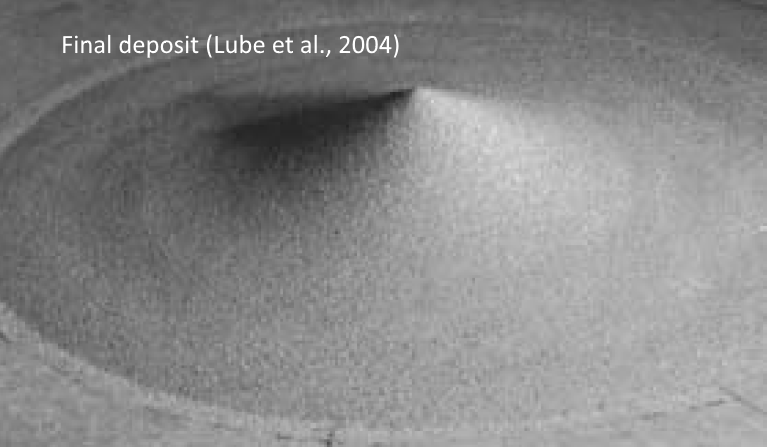} \\
  \end{tabular}
  \caption{The evolution of granular material column for different initial
  aspect ratios: $a=0.55$ and $2.75$; the SPH results compared with 
  the experimental photographs by 
  \emph{Lube et al. (2004)}~\cite{Lube et al. 2004}.}
  \label{fig:shapes-3d}
\end{figure}

For the 3D model, it was decided to compare the SPH results with the experimental data 
by \emph{Lube et al. (2004)}~\cite{Lube et al. 2004}. For validation purposes we have chosen two different aspect ratios: $a=0.55$ and $a=2.75$.
The results are presented in Fig.~\ref{fig:shapes-3d}.
The performed simulations gave very realistic results.
Using the SPH approach, we were able to reproduce the typical for $a<0.74$,
see~\cite{Lube et al. 2004}, circular discontinuity on the surface 
of the column which separates an outer (slumping) region from a non-deformed
inner part of the deposit.
For the column of $a=2.75$ the calculated shape evolution of the column
agrees very well with the experimental data presented in detail in~\cite{Lube et al. 2004}
 (Figure~4).

\subsection{Run-out distances} \label{sec:run-out distances}
One of the most relevant validation criteria for the numerical model is to try to reproduce
the scaling laws for the run-out distance. The experimental data for the three-dimensional
column was first proposed by \emph{Lube et al. (2004)}~\cite{Lube et al. 2004}:
\begin{equation}
  \frac{r_\infty - r_0}{r_0} \simeq 
  \begin{cases}
    1.24 a,      &  a < 1.7, \\
    1.6 a^{1/2}, &  a \geq 1.7. \\
  \end{cases}
\end{equation}
Slightly different results were obtained by \emph{Lajeunesse et al. (2005)}~\cite{Lajeunesse et al. 2005} for semi-circular (half of column) initial configuration:
\begin{equation}
  \frac{r_\infty - r_0}{r_0} \simeq 
  \begin{cases}
    a,       &  a < 3, \\
    a^{1/2}, &  a \geq 3. \\
  \end{cases}
\end{equation}

Many more empirical and numerical experiments were performed for granular collapses of the two-dimensional columns. \emph{Lube et al. (2005)}~\cite{Lube et al. 2005} were releasing 
a granular columns confined between two vertical walls. The authors obtained the following scaling:
\begin{equation}
  \frac{r_\infty - r_0}{r_0} \simeq 
  \begin{cases}
    1.2 a,       &  a < 2.3, \\
    1.9 a^{2/3}, &  a \geq 2.3. \\
  \end{cases}
\end{equation}
Independently \emph{Lajeunesse et al. (2005)} obtained a similar result:
\begin{equation}
  \frac{r_\infty - r_0}{r_0} \simeq 
  \begin{cases}
    a,       &  a < 3, \\
    a^{2/3}, &  a \geq 3. \\
  \end{cases}
\end{equation}
The two-dimensional numerical experiments were performed by 
\emph{Staron and Hinch (2005)}~\cite{Staron and Hinch 2005} who obtained the relation:
\begin{equation}
  \frac{r_\infty - r_0}{r_0} \simeq 
  \begin{cases}
    2.5 a,       &  a < 2, \\
    3.25 a^{0.7}, & a \geq 2. \\
  \end{cases}
\end{equation}
Many other authors including \emph{Zenit (2005)}~\cite{Zenit 2005}, 
\emph{Utili et al. (2015)}~\cite{Utili et al. 2015} who used the DEM approach
and \emph{Crosta et al. (2009)}~\cite{Crosta et al. 2009} who performed FEM simulations
obtained fairly consistent results for the run-out distance.

\begin{figure}
  \centering
  \includegraphics[width=0.7\textwidth]{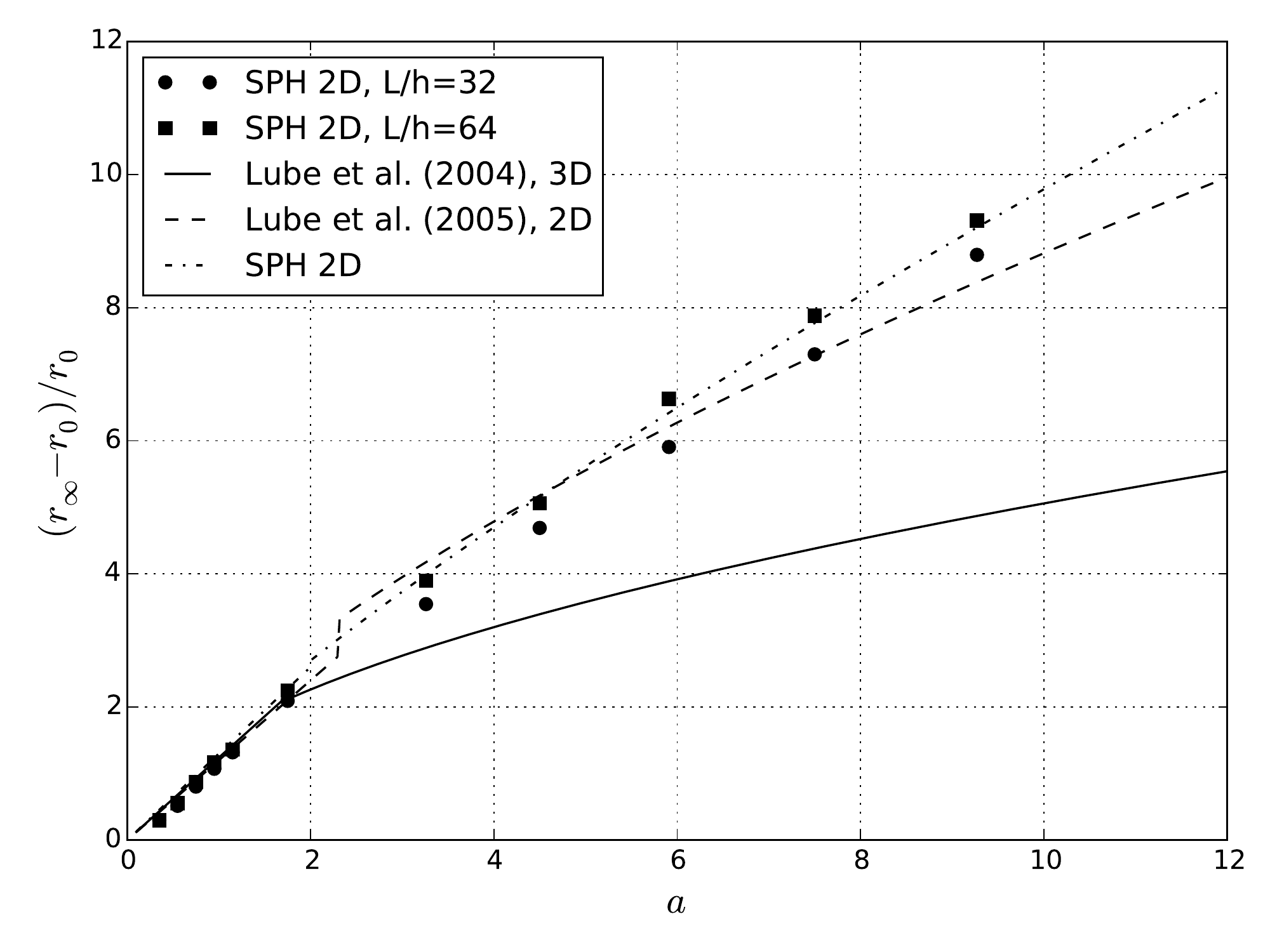}
  \caption{The non-dimensional incremental run-out distance as a function of the aspect ratio for two-dimensional results; the SPH solution compared with two-~\cite{Lube et al. 2005} and three-dimensional~\cite{Lube et al. 2004} experimental data.}
  \label{fig:range-2d}
\end{figure}

\begin{figure}
  \centering
  \includegraphics[width=0.7\textwidth]{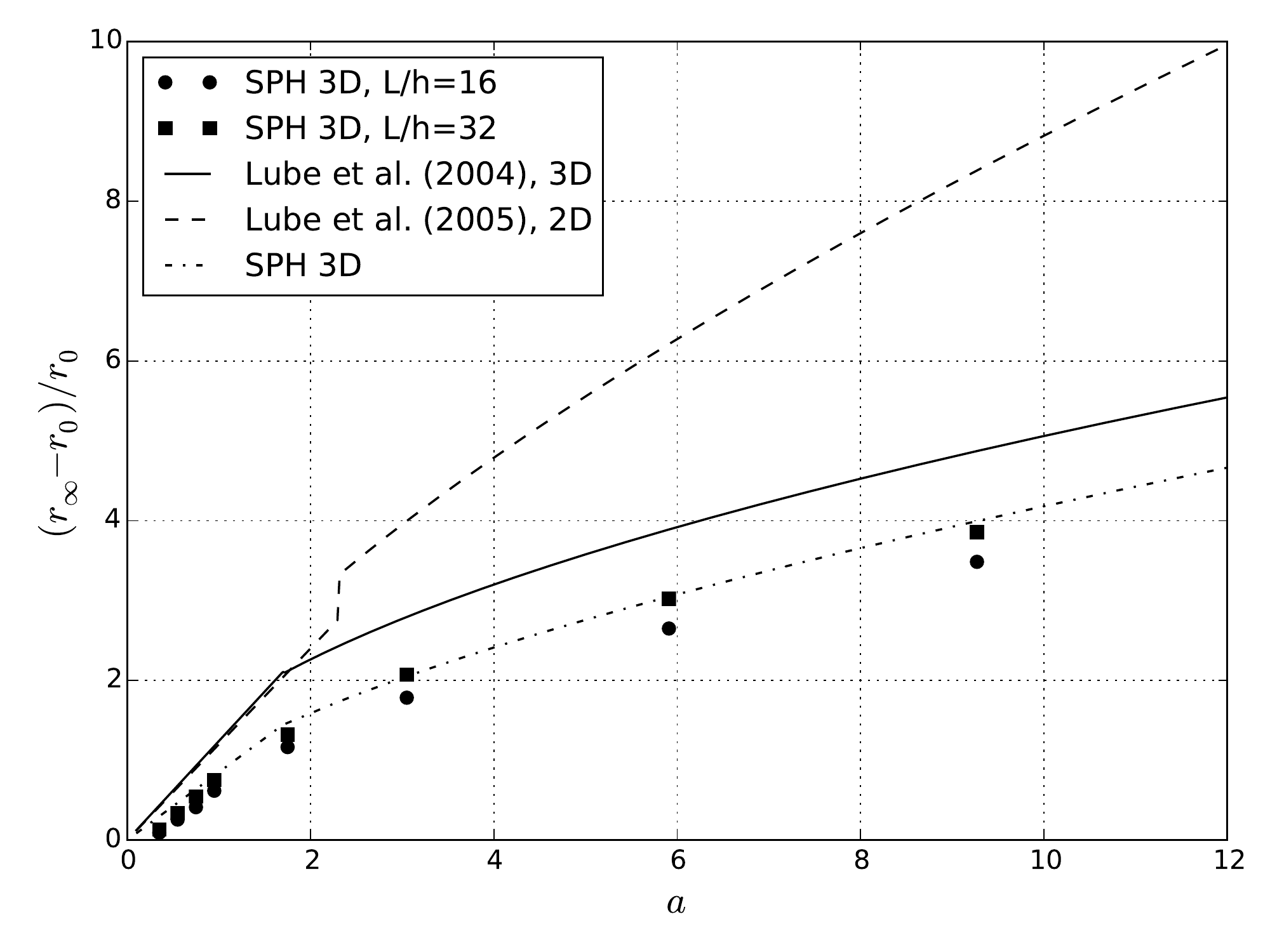}
  \caption{The non-dimensional incremental run-out distance as a function of the aspect ratio for three-dimensional results; the SPH solution compared with two-~\cite{Lube et al. 2005} and three-dimensional~\cite{Lube et al. 2004} experimental data.}
  \label{fig:range-3d}
\end{figure}

The obtained results of the relation between the normalized run-out distance and 
the initial column aspect ratio for 2D and 3D cases are respectively presented in Figs.~\ref{fig:range-2d} and \ref{fig:range-3d}.
As the reference data we decided to 
plot two- and three-dimensional experimental solution obtained in 
\emph{Lube et al.}~\cite{Lube et al. 2005,Lube et al. 2004}.

Here we show that in the case of 2D columns 
the normalized run-out distances for higher aspect ratios 
are slightly overestimated 
when compared with the experimental data~\cite{Lube et al. 2005}.
In the present SPH approach, we obtained:
\begin{equation}
  \frac{r_\infty - r_0}{r_0} \simeq 
  \begin{cases}
    1.3 a,       &  a < 2, \\
    1.55 a^{4/5}, & a \geq 2. \\
  \end{cases}
\end{equation}
However, in general, the results show a good agreement with the reference data.

In the case of 3D columns, we obtained somewhat less accurate results compared with
the reference data~\cite{Lube et al. 2004}. The SPH results appear to be underestimated.
One reason of such a behavior is smaller numerical resolution compared 
with the 2D simulations. The obtained scaling law is:
\begin{equation}
  \frac{r_\infty - r_0}{r_0} \simeq 
  \begin{cases}
    0.85 a,       &  a < 1.7, \\
    1.05 a^{3/5}, & a \geq 1.7. \\
  \end{cases}
\end{equation}

\subsection{Energy contribution} \label{sec:energy contribution}

The potential energy of the column at any time is
\begin{equation}
  E_p = \sum_a m_a g h_a,
\end{equation}
where $h_a$ is the height of the particle $a$.
The column kinetic energy can be calculated from
\begin{equation}
  E_k = \frac{1}{2} \sum_a m_a u_a^2.
\end{equation}
Due to the viscosity, or (in the scale of a single grain) 
non-elastic collisions between grains,
a part of the potential energy does not transform into the kinetic energy, 
but it gets dissipated
\begin{equation}
  E_{dis}(t) = E_p(0) - E_p(t) - E_k(t).
\end{equation}

\begin{figure}
  \centering
  \includegraphics[width=0.49\textwidth]{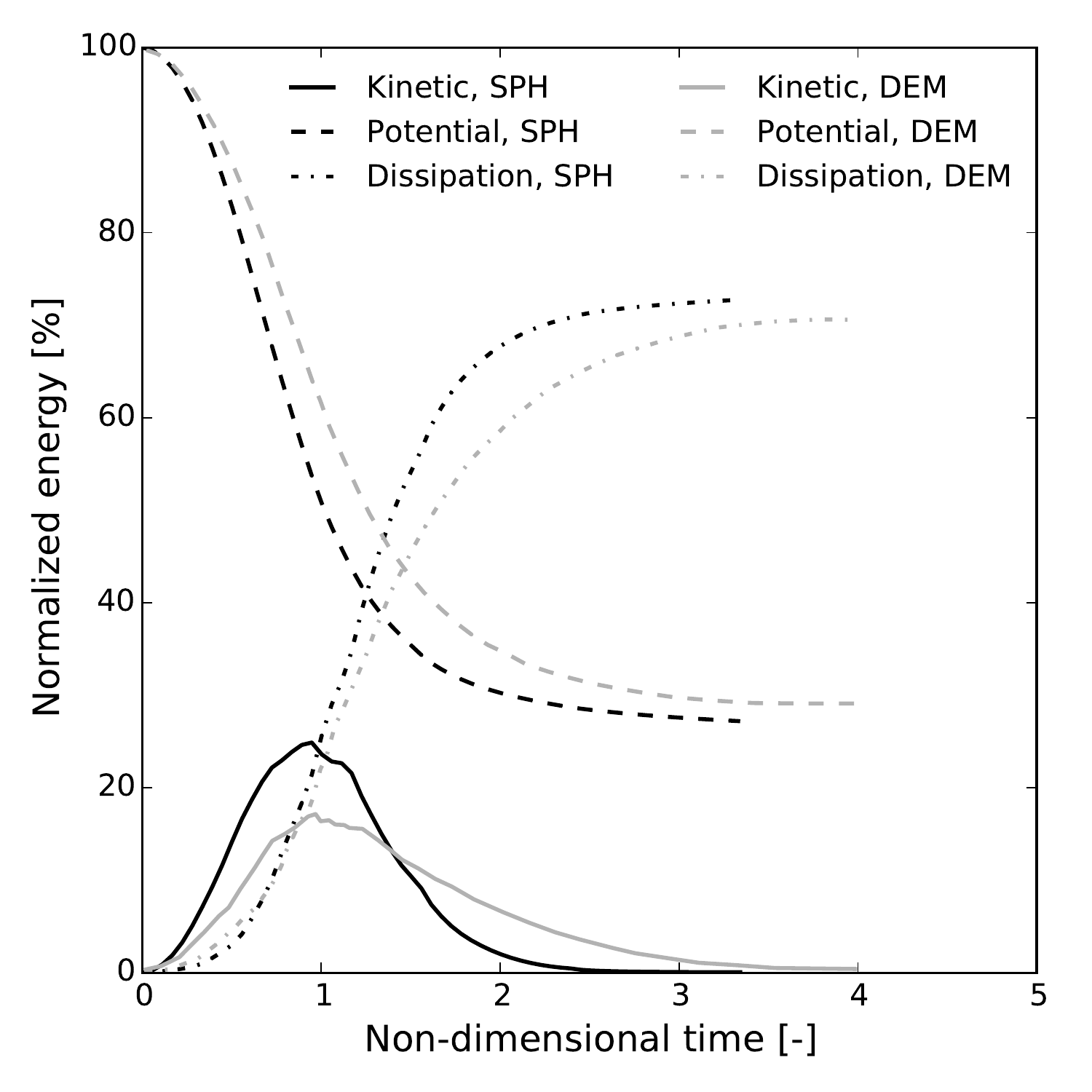}
  \includegraphics[width=0.49\textwidth]{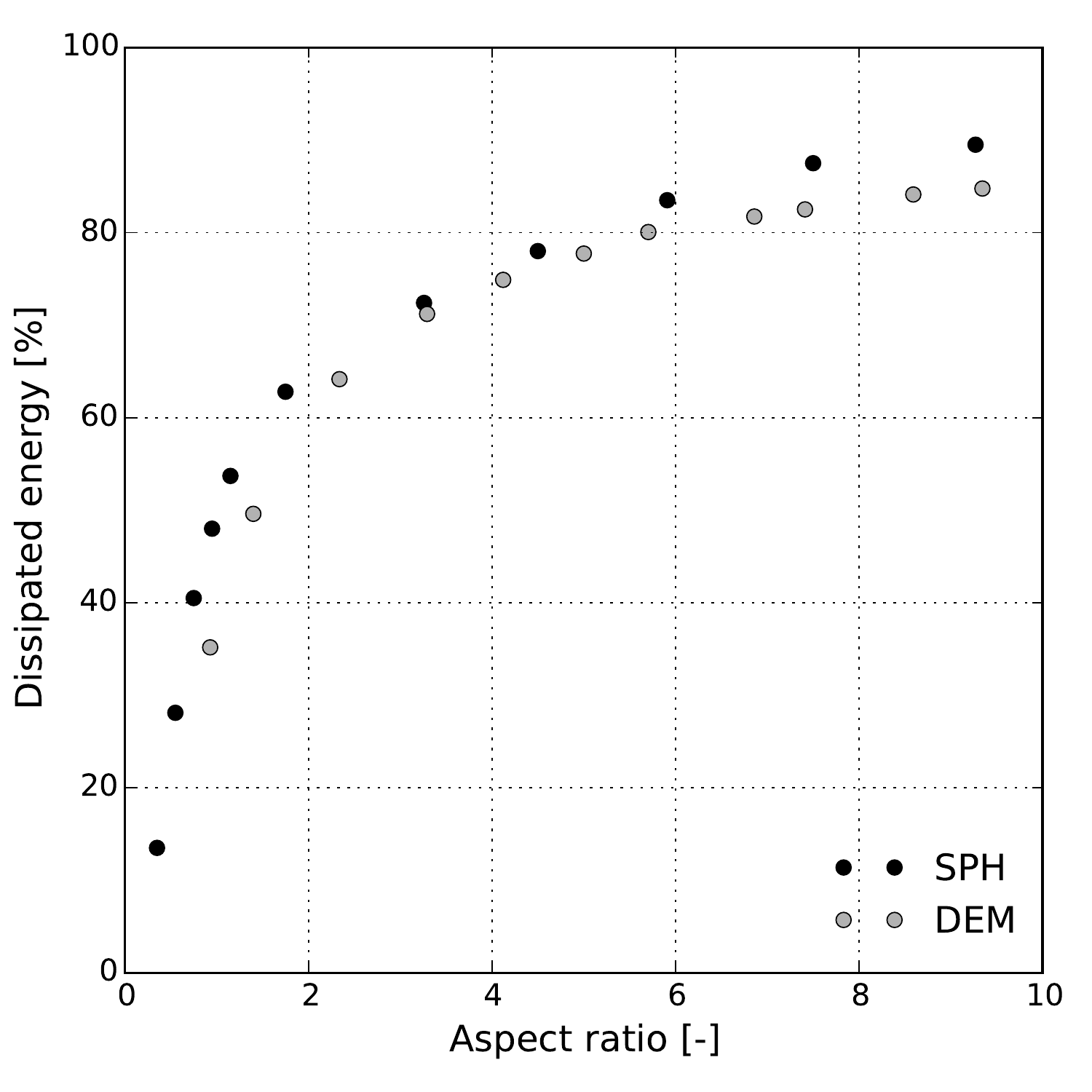}
  \caption{(left) Energy evolution for the granular column of the aspect ratio $a=3.26$;
  (right) total energy dissipated during the flow as a function of the aspect ratio;
  energy normalized by the initial potential energy;
  the SPH result compared with the DEM reference data~\cite{Utili et al. 2015}.}
  \label{fig:energy}
\end{figure}

At the beginning, the entire energy is accumulated as the potential energy.
As time passes, particles start to fall downwards with the potential energy
being transformed into the kinetic energy and some heat (dissipation).
Figure~\ref{fig:energy}(left) shows the energy evolution of the granular column for the aspect
ratio $a=3.26$. In this plot the SPH results are compared with the DEM simulations obtained in~\cite{Utili et al. 2015}.
In both models the kinetic energy exhibits a peak at about $t/T = 1$.
The obtained results show small discrepancies between SPH and DEM, 
however these differences can be minimized
adjusting the minimal and maximal viscosity in the considered rheological model.

In order to validate the SPH approach for different values of the aspect ratio,
we decided to compare the total dissipated energy for different values of $a$
calculated using the SPH method with the DEM simulations~\cite{Utili et al. 2015}.
The obtained results are presented in Fig.~\ref{fig:energy}(right).
Both SPH and DEM models give similar relation between the dissipated energy 
and aspect ratio. Small overestimation of the SPH results is also observed 
in Fig.~\ref{fig:energy}(left).

\subsection{Inclination of the failure plane} \label{sec:inclination of failure plane}

\begin{figure}
  \centering
  \includegraphics[width=0.5\textwidth]{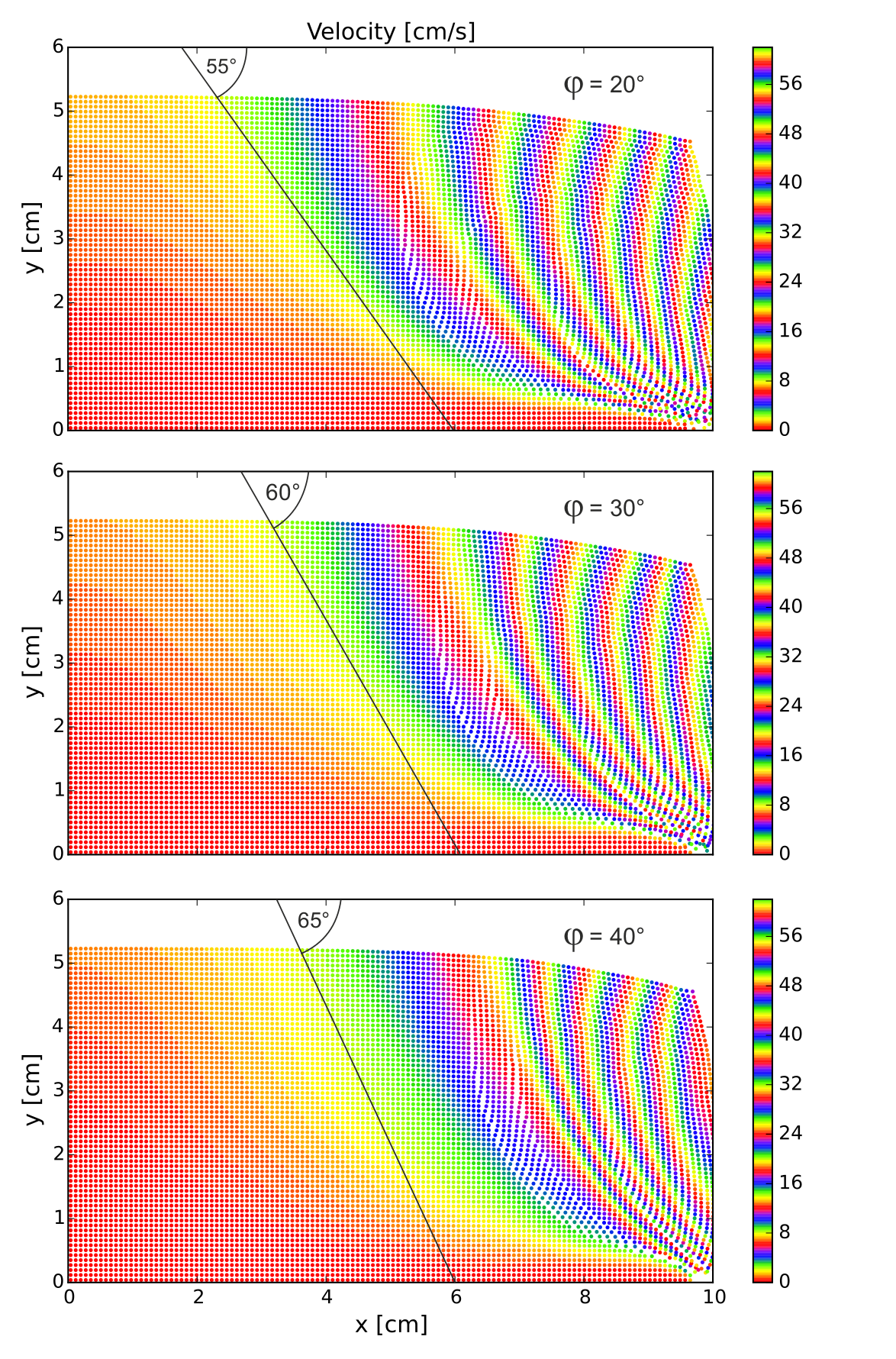}
  \caption{Active failure state of the granular sample; 
  the results obtained for three different internal friction angles: (a) $20^\circ$, (b) $30^\circ$ and (c) $40^\circ$.}
  \label{fig:angle}
\end{figure}

According to the Rankine's theory of earth pressure~\cite{Rankine 1857}, 
the inclination of the failure plane to the horizontal, $\theta_f$
can be approximated as:
\begin{equation} \label{rankine}
  \theta_f = 45^{\circ} + \frac{\varphi}{2}.
\end{equation}
In order to check whether the SPH approach can correctly predict the relation (\ref{rankine}),
we decided to perform three simulations of granular column collapse with 
different values of the internal friction angle $\varphi=20^{\circ}$, $30^{\circ}$, and $40^{\circ}$.
It is important to note that values $20^{\circ}$ and $40^{\circ}$ correspond 
to the extreme values observed in nature.
To visualize the inclination angle of the slope failure plane, we decided to plot 
the velocity fields using the spectral color map, see Fig.~\ref{fig:angle}.
The presented results were obtained for the collapsing column of aspect ratio $a=0.55$
at $t = 40$ ms. 
The obtained failure angles agree well with the relation (\ref{rankine}). 
Similar test performed for different values of aspect ratios show no vital differences
in relation between $\theta_f$ and $\varphi$. The only problem that must be noted is
small systematic decline of the $\theta_f$ values for vary late time steps. 
This behavior is due to decrease of the height of the deposit caused by 
lowered $\mu_{\text{solid}}$ (compared with reality) due to the numerical efficiency, see Section~(\ref{sec:introduction}).

\subsection{Numerical efficiency} \label{sec:numerical efficiency}

\begin{figure}
  \centering
  \includegraphics[width=0.49\textwidth]{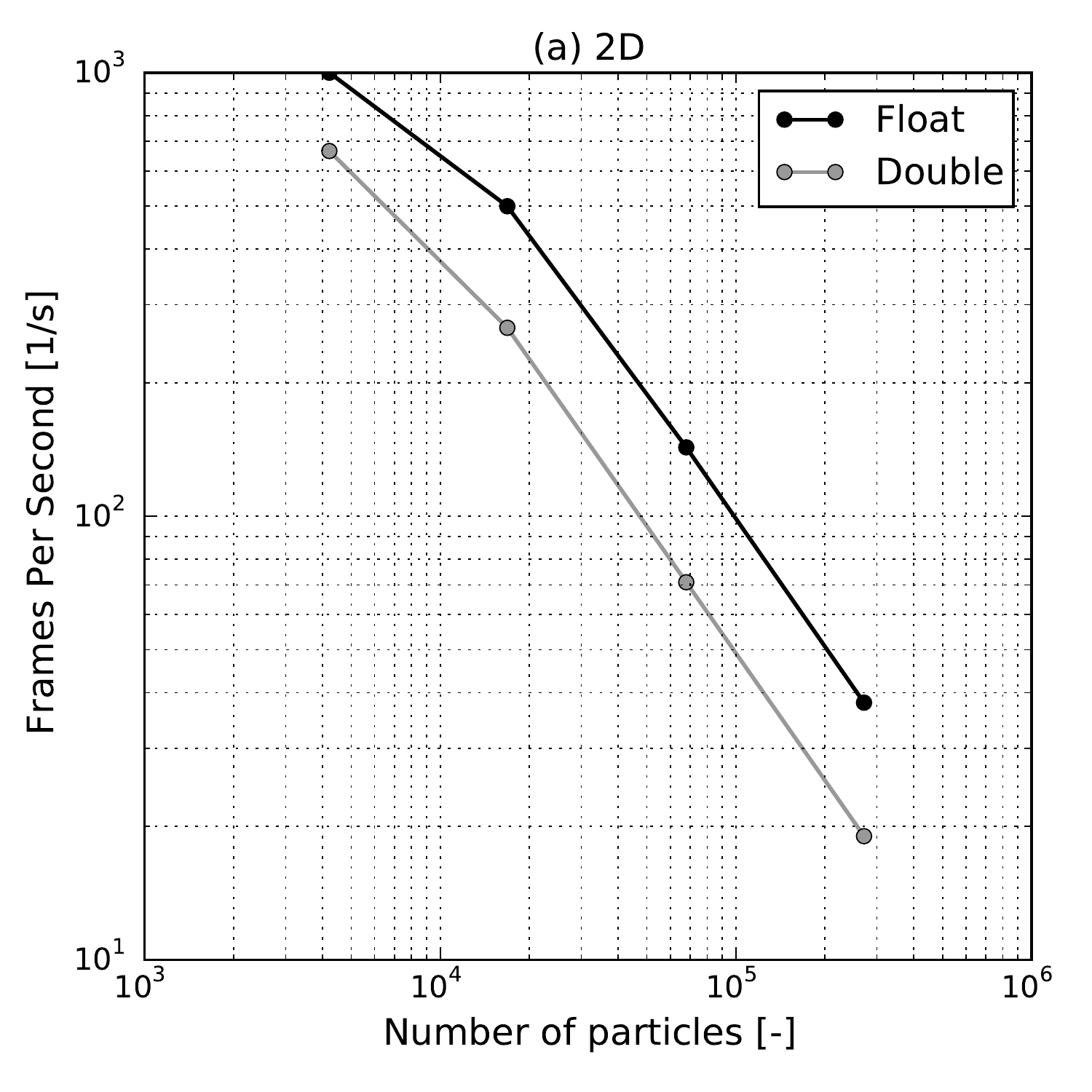}
  \includegraphics[width=0.49\textwidth]{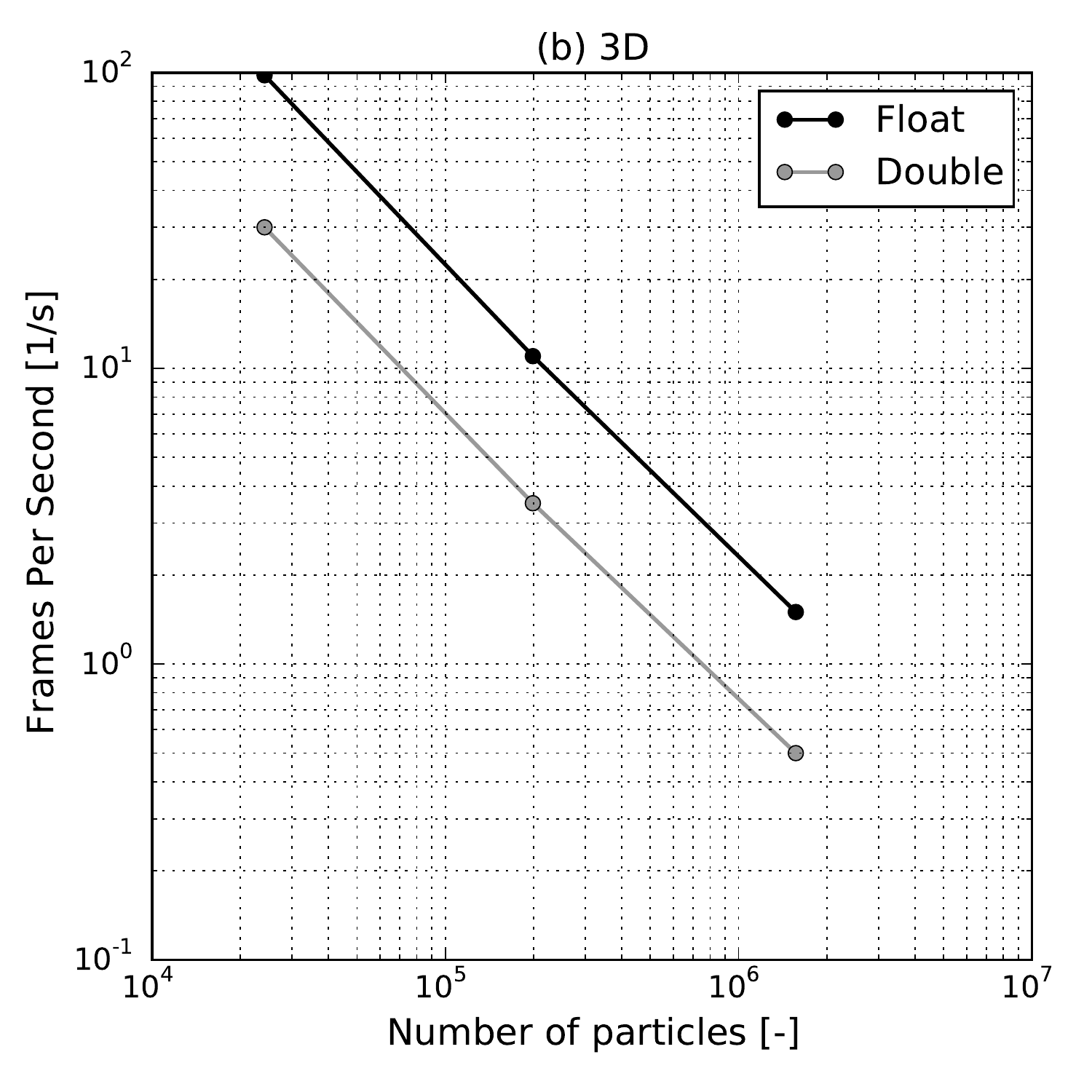}
  \caption{Number of frames per second as a function of the SPH particles in the domain;
  the initial aspect ratio is $a=0.55$;
  (a) two-dimensional results for $h/\Delta r = 2$, (b) three-dimensional results
  for $h/\Delta r = 1.5625$.}
  \label{fig:performance}
\end{figure}

For the performance analysis we decided to use the \emph{Nvidia GeForce GTX980}
GPU ($2048$ cores, $1126$ MHz clock, $4$ GB of memory).
Figure~\ref{fig:performance} presents the measured 
Frames (time steps) Per Second (FPS) as 
a function of the number of particles in the domain. In 2D, the use of the double
precision floating numbers decreases the computational time twice.
In single precision the calculations took from about $2$ min for about $N=4 \cdot 10^3$
particles up to about $1$ h for $N=2.5\cdot 10^5$.
In 3D, the use of the single precision numbers increases performance by factor $3$.
For the lowest used resolution ($N=2.5 \cdot 10^4$) simulations took $23$ min,
while for the highest resolution ($N=1.5 \cdot 10^6$) about $26$ h (single precision).
It is important to note that the used GPU is based on the \emph{Maxwell} micro-architecture
in which the double precision performance is 1/32 of the single precision performance.
Therefore, the main reason of observed decrease of the 
performance with double precision in 3D
(compared with the single precision) is much larger number of interactions 
under the kernel hat (larger number of simple math operations).
The difference between the double and single precision performance should be much
smaller using the \emph{Kepler} micro-architecture 
(\emph{Nvidia Tesla} and \emph{Nvidia GeForce 700 series}) where the double precision computations are only 4 times less efficient. The numerical efficiency can be
further improved by using more than one GPU.

\section{Conclusions} \label{sec:conclusions}

In the present work, the ability to model granular materials using the SPH method
and the visco-plastic model has been studied. For this purpose, a set of numerical 
calculations (in 2D and 3D) of the fundamental problem of the collapse of initially vertical cylinders of granular materials has been performed. 
In order to validate the proposed model, the granular deposit evolution, 
the run-out distances, the energy contribution and the inclination
of the failure plane were compared with the analytical, experimental
and other numerical data. The obtained results showed good agreement 
with the reference data. All the inaccuracies that we observed during simulations
were caused mainly by two factors: not perfectly matched parameters of the
used rheological model or too low numerical resolution (limited hardware resources).
In order to reduce the effect of particle clustering (the tensile instability)
-- the problem signalized in \cite{Bui et al. 2008}, we decided to suitably
choose the kernel function which significantly reduces this problem.
In fact, the tensile instability was not observed in the obtained results.
Taking advantage of GPU efficiency, it was possible to run computationally heavy
simulations on the cheap desktop computer. The performed analysis showed that 
the single precision of floats is not enough to correctly perform simulations
with the used rheological model. The double precision calculations increase
the computational effort on GPU, but, the obtained numerical efficiency 
is still very high.
It is important to note here that for dry granular materials the methods such as
the DEM allow for much more accurate and efficient calculations. However, when we
consider complex debris flow constituted of rocks and mud, for which it may be difficult
to define the interaction between solid particles, the continuum methods, such as the
introduced SPH, appear to be much more useful. Another advantage of the SPH approach is
its ability to model complex multi-phase flows involving eg. fluid-granular phase interactions. 
This work is an intermediate step in a complete project which aims at
simulating the interaction of sea waves and currents with a seabed. 
The satisfactory results obtained for the dynamics of dry sand
with the simple rheological model are an encouragement to pursue
along that direction.

\section*{Funding}
This study was funded by the National Science Centre, Poland via 
grant Opus 6 no DEC-2013/11/B/ST8/03818.

\section*{Conflict of Interest}
The author declare that he has no conflict of interest.

%
%

\end{document}